\documentclass[aps,prl,twocolumn]{revtex4}

\usepackage{graphicx}
\usepackage{amssymb}
\usepackage{latexsym,amsmath,amsthm}
\usepackage[dvips]{color}
\begin{document}

\title{Incoherent matter-wave solitons: Mutual self-trapping of
\\ a Bose-Einstein condensate and its surrounding thermal cloud}

\author{H. Buljan,$^{1,2}$ M. Segev,$^{1}$ and A. Vardi$^{3}$}
\affiliation{ $^{1}$Physics Department, Technion -
Israel Institute of Technology, Haifa 32000, Israel}
\affiliation{
$^2$Department of Physics, University of Zagreb, PP 332, Zagreb, Croatia}
\affiliation{
$^3$Department of Chemistry, Ben Gurion University of Negev, Beer Sheva, Israel}
\date{\today}

\begin{abstract}
We show that a Bose-Einstein condensate and a portion of
its surrounding thermal cloud can exhibit mutual self-trapping,
supported by the attractive particle interactions, and
not by the external confinement. This type of dynamics is
characteristic of composite random-phase solitons.
\end{abstract}


\maketitle


The physics of quantum-degenerate, interacting bose gases closely resembles
the behavior of  light in nonlinear media. The dynamics of Bose-Einstein
condensate (BEC) at zero-temperature
is within the Gross-Pitaevskii (GP) mean-field theory
described by the nonlinear Schr\"odinger equation (NLSE)
for the condensate order parameter. The same equation describes the evolution
of coherent light in nonlinear Kerr medium.
This analogy has opened the way for the field of nonlinear atom optics
\cite{Lens,Rolston} with striking demonstrations of familiar nonlinear
optics phenomena such as four wave mixing \cite{Deng}, superradiant
Rayleigh scattering \cite{Inouye1}, and matter-wave amplification
\cite{Kozuma,Inouye2}, carried out with matter-waves.
One such phenomenon is the formation of matter-wave solitons \cite{Ruprecht,PerezG,Burger,Denschlag,Busch,Smerzi,Khajkovic,Strecker,Salasnich,Eiermann}.
Experimentally, dark solitons \cite{Burger,Denschlag} and bright gap solitons
\cite{Eiermann} were observed in BECs with repulsive interactions,
whereas bright solitons  \cite{Khajkovic,Strecker}
were demonstrated in systems with attractive interactions.
These experimental results are augmented by extensive theoretical work
including predictions on bright \cite{Ruprecht,PerezG} and
dark \cite{Busch} matter-wave solitons, lattice solitons \cite{Smerzi},
and soliton trains \cite{Salasnich}.
To the best of our knowledge, all previous theoretical efforts
on matter-wave solitons have utilized the zero-temperature GP
mean-field theory. However, in a realistic system, elementary excitations
arising from thermal and/or quantum fluctuations are always present, and
the BEC dynamics may be considerably affected by the
motion of excited atoms around it (thermal cloud), giving rise
to new nonlinear matter-wave phenomena.

Here we present an example of such novel phenomena,
and show that a BEC and a portion of its surrounding thermal
cloud can exhibit mutually self-trapped motion.
This motion is achieved via attractive interactions
between particles, and not by the external confinement.
We emphasize that the finite-temperature self-trapping produces a
truly novel type of matter-wave solitons, where localization is
attained not only in spatial density but also in spatial correlations.
These self-trapped incoherent matter-waves are analogous to
composite random-phase (incoherent) optical solitons \cite{Mitchell,Equ,Mitchell1};
one component is the BEC while the other is a part of its thermal cloud.
An important result of this work is that the established
analogy between zero-temperature BECs and coherent nonlinear optics
can be elevated to the analogy of incoherent light
behavior in nonlinear media and BECs at finite-temperatures.

The incoherent matter-wave self-trapping is illustrated within
the Hartree-Fock (HF) approximation. First, we solve the static
HF problem \cite{Leggett,Griffin,Stoof_HF} describing Bose gas
with attractive interactions in a confining harmonic potential.
We consider the range of (higher) temperatures where
excitations are mostly particle-like, and HF solutions are approximately
equal to the solutions obtained using the Hartree-Fock-Bogoliubov (HFB)
theory in the Popov approximation \cite{Griffin}.
We focus on solutions with significant population of the
first-excited state. In order to establish mutual-self-trapping,
we turn-off the external harmonic confinement, and evolve the BEC and
the thermal cloud within the time-dependent HF theory (TDHF)
\cite{Proukakis1,Minguzzi,Mukamel}. The signature of mutual self-trapping
is a slow separation of the total density profile into two humps, which propagate
almost in parallel, and are then pulled back to almost
recover the initial density profile.
This self-trapped motion is compared to the evolution of precisely
the same initial state when both the trap and the interparticle interactions
are turned off simultaneously, exhibiting fast matter-wave
dispersion.

We consider a system of $N$ interacting bosons placed in a quasi
one-dimensional (Q1D) cigar-shaped harmonic potential
$V_{ext}(x,y,z)=(\omega_x x^2+\omega_{\perp} y^2+\omega_{\perp} z^2)/2$,
where $\omega_{\perp}\gg\omega_x$ denote the transverse and the
longitudinal frequencies of the trap, respectively.
The interparticle interaction is approximated by the Q1D contact potential
$V(x_1-x_2)=g_{1D}\delta(x_1-x_2)$, where $g_{1D}=-2\hbar^2/ma_{1D}$,
$a_{1D}\approx -a_{\perp}^2/a_{3D}$ is the effective 1D scattering length \cite{Dunjko,Moritz},
$m$ is the particle mass, $a_{\perp}=\sqrt{\hbar/m\omega_{\perp}}$ is the size
of the lowest transverse mode, while $a_{3D}$ is the 3D scattering length.
At finite temperatures, this system can be described by the
Hartree-Fock Bogoliubov (HFB) pairing theory \cite{Griffin}.
We are interested in the regime of higher temperatures, where
the thermal cloud is sufficiently large; in this limit,
anomalous pairings are negligible, the excitations are mostly particle-like,
and the HFB formalism can be approximated with the HF theory \cite{Griffin}.
In our case, the system is initially at thermal equilibrium.
The density of the condensate $|\langle \hat \psi(x,t) \rangle|^2$,
and the noncondensed particles $\langle \tilde \psi^{\dagger}(x,t)\tilde \psi(x,t)\rangle$
are calculated within the static HF approximation
\cite{Leggett,Griffin,Stoof_HF}:

\begin{equation}
H_{sp}\phi_0^{(s)}+g_{1D}[n_c^{(s)}(x)+2 n_t^{(s)}(x)]\phi_0^{(s)}
=e_0\phi_0^{(s)}(x),
\label{staticcon}
\end{equation}
\begin{equation}
H_{sp}u_j^{(s)}+g_{1D}[2n_c^{(s)}(x)+2n_t^{(s)}(x)]u_j^{(s)}
=e_ju_j^{(s)}(x).
\label{staticu}
\end{equation}
Here, $H_{sp}=-\frac{\hbar^2}{2m}\frac{\partial^2}{\partial x^2}
+\frac{1}{2}m\omega_x^2 x^2$;
$\phi_0^{(s)}(x)$ is the condensate wavefunction, with eigenvalue $e_0$;
$u_j^{(s)}(x)$ are the wavefunctions of the excited states, with
eigenvalues $e_j$, $j=1,2,\ldots$;
$n_c^{(s)}(x)=N_c|\phi_0^{(s)}(x)|^2$ is the condensate density,
where $N_c$ is the number of condensed particles;
$n_t^{(s)}(x)=\sum_j N_j |u_j^{(s)}|^2$ is the density of the
excited particles; the number of particles populating the $j$th
excited state, $N_j$, is determined by the Bose
distribution, $N_j=[\exp(\frac{e_j-\mu}{k_BT})-1]^{-1}$.
The chemical potential $\mu$ is set by the constraint
$N=N_c+\sum_jN_j$; all wavefunctions are normalized to unity.
We solve Eqs. (\ref{staticcon}) and (\ref{staticu}) self-consistently,
and use the static solution as the initial condition
to study dynamics without confinement.

The dynamics is studied within the TDHF approximation, which involves
the coupled evolution of the condensate wavefunction
$\phi_0(x,t)=\langle \hat \psi(x,t) \rangle$, and the correlation function
$\rho(x_1,x_2,t)=\langle \tilde\psi^{\dagger} (x_2,t) \tilde\psi (x_1,t) \rangle$
\cite{Proukakis1,Minguzzi,Mukamel}:

\begin{equation}
i\hbar \frac{\partial \phi_0(x,t)}{\partial t}=
H_{sp}\phi_0+g_{1D}[n_c(x,t)+2 n_t(x,t)]\phi_0
,
\label{con}
\end{equation}
\begin{eqnarray}
i\hbar \frac{\partial \rho}{\partial t}  & = &
[H_{sp}(1)-H_{sp}(2)]\rho
 +  2g_{1D}[ n_c(x_1,t) + n_t(x_1,t) \nonumber \\
& - & n_c(x_2,t)- n_t(x_2,t)]\rho(x_1,x_2,t),
\label{exctrho}
\end{eqnarray}
where $n_c(x,t)=N_c|\phi_0(x,t)|^2$ and $n_t(x,t)=\rho(x,x,t)$.
The correlation function $\rho$ is a part of the single particle
density matrix $\langle \hat\psi^{\dagger}(x_2,t)\hat\psi(x_1,t) \rangle
=\phi_0^*(x_2,t)\phi_0(x_1,t)+\rho(x_1,x_2,t)$
corresponding to excitations. Equations of motion
(\ref{con}) and (\ref{exctrho}) for the condensate wavefunction and the
excitations describe the evolution of the BEC and the thermal cloud
even outside of equilibrium \cite{Proukakis1}.
Initial conditions at $t=0$ are $\phi_0(x,t=0)=\phi_0^{(s)}(x)$ and
$\rho(x_1,x_2,t=0)=\sum_jN_j u^{*(s)}(x_2)u_j^{(s)}(x_1)$,
and the evolution is performed without the external potential.
However, rather than to use
Eqs. (\ref{con}) and (\ref{exctrho}), we shall use a fully equivalent,
but numerically more convenient approach. The solution
for $\rho(x_1,x_2,t)$ may be constructed from
$\rho(x_1,x_2,t)=\sum_jN_j u_j^*(x_2,t)u_j(x_1,t)$, where
functions $u_j(x,t)$ evolve according to an infinite set
of coupled equations,

\begin{equation}
i\hbar \frac{\partial u_j(x,t)}{\partial t}  =
H_{sp}u_j +g_{1D} 2 [ n_c(x,t)+ n_t(x,t)]u_j,
\label{exctu}
\end{equation}
where $n_t(x,t)=\sum_j N_j |u_j(x,t)|^2$.
Note that when the time-dependence of the wavefunctions
is simply $\phi_0(x,t)=\phi^{(s)}(x)\exp(-ie_0 t/\hbar)$
and $u_j(x,t)=u_j^{(s)}(x)\exp(-ie_jt/\hbar)$, the equations of motion
(\ref{con}) and (\ref{exctu}) reduce to the static
HF equations (\ref{staticcon}) and (\ref{staticu}).

\begin{figure}
\centering
\includegraphics[scale=0.5]{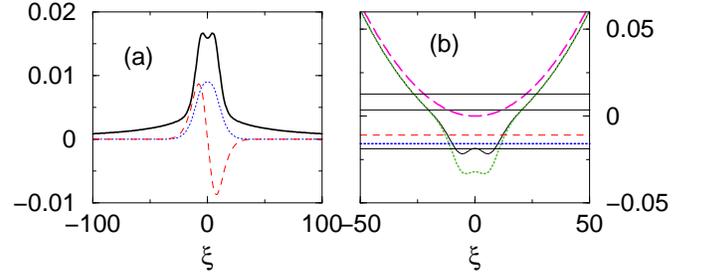}
\caption{(color online) Solution of the static Hartree-Fock
equations. (a) The total density (solid line), the bell-shaped
condensate wavefunction (dotted line), and the dipole-shaped
first excited state (dashed line). (b) The (mean-field) potential
seen by the condensate (solid line), by the non-condensed
particles (dotted line), and the external harmonic potential
without the mean field (dashed line). Horizontal lines
from the lowest one up depict the chemical potential $\mu$,
the condensate eigenvalue $e_0$, eigenvalue $e_1$ of the
dipole-mode, and eigenvalues $e_2$, and $e_3$, expressed in units of
$\hbar \omega_{\perp}$.
}
\label{fig1}
\end{figure}

In what follows we present results of a numerical calculation based on the
described HF formalism, demonstrating the mutually self-trapped
motion of the BEC and a portion of its thermal cloud.
The parameters of the calculation are chosen to
resemble the experimental parameters of Ref. \cite{Khajkovic}. We
consider $N=2.2\ 10^4$ $^{7}$Li atoms in a harmonic trap with
$\omega_{\perp}=4907$ Hz ($a_{\perp}=\sqrt{\hbar/m\omega_{\perp}}\approx 1.35\ \mu$m),
and $\omega_x=35$ Hz ($a_{x}=\sqrt{\hbar/m\omega_{x}}\approx 16.0\ \mu$m).
The 3D scattering length $a_{3D}=-3.1\ 10^{-11}$ m corresponds to a
nonlinear parameter of $N|a_{3D}|\approx 0.68 \ \mu$m, and is tunable by the
Feshbach resonance technique \cite{Khajkovic}. The temperature is
$k_BT/\hbar\omega_{\perp}=16$. The notable differences from
the experiment of Ref. \cite{Khajkovic} are in the longitudinal
trapping frequency, which is an order of magnitude smaller in our
calculation, and in the temperature, which is here chosen
to produce a sufficiently large thermal cloud and
to ensure the validity of the HF formalism.

While the temperature $k_BT$ in our simulation is higher than the transverse level
spacing $\hbar \omega_{\perp}$ whereas a 'true' 1D geometry calls for
$k_BT<\hbar \omega_{\perp}$ \cite{Dunjko,Moritz},
the use of a Q1D formalism is still justified because the first
$\omega_x/\omega_{\perp}\sim 140$ states are essentially 1D
(they are in the lowest state of the transverse Hamiltonian).
As shown below,
only a few of the lowest excited states actively participate
in the self-trapping process. Therefore, a proper inclusion of the
transverse dimension in the calculation would lead to some rescaling
of the parameters, but would not influence the self-trapping process
observed in our quasi-1D calculation. Moreover, the simulations as well
as the experiment of \cite{Khajkovic}, are far from the Tonks-Girardeau
regime of impenetrable bosons \cite{Dunjko};
our calculations are all in the weak interaction regime
$N|a_{1D}|/a_x\sim 10^7\gg 1$, thus justifying the use of a mean-field
approach (we note parenthetically that
the weak interaction regime for a 1D gas is attained
'counterintuitively' at {\it high} densities \cite{Dunjko}).
The stability of the confined, attractively interacting condensate
against collapse \cite{Ruprecht,Kagan1} was numerically verified
by evolving it with random initial noise on top of all
equilibrium modes; the stability is underpinned by the use of
parameters resembling the experiment \cite{Khajkovic}.

The system is initially in equilibrium.
Fig. \ref{fig1}(a) illustrates the total density
$n_c^{(s)}(x)+n_t^{(s)}(x)$ of the stationary HF calculation.
The total density profile (solid line) is double humped, offering
a clear signature of the significant population of the
first excited $u^{(s)}_1$ state, which has a dipole-like
spatial profile (dashed line). The condensate fraction is
$N_c/N\approx 0.24$, while the population of the first excited state
is $N_1/N\approx 0.093$. The excitation energies and the
mean-field potentials affecting the condensate and
the non-condensed cloud are plotted in Fig. \ref{fig1}(b).
The eigenvalue $e_1$ of the first excited state is
below zero, along with the condensate eigenvalue $e_0$ and the
chemical potential $\mu$ ($\mu\rightarrow e_0$ when $T\rightarrow 0$).
Thus, a large fraction of the thermal cloud [$N_1/(N-N_c)\approx 0.12$]
is self-localized jointly with the condensate, rather than localized by
the external harmonic potential. The same conclusion is also inferred from
the double-well structure of the mean-field potentials [Fig. \ref{fig1}(b)].

\begin{figure}
\centering
\includegraphics[scale=0.5]{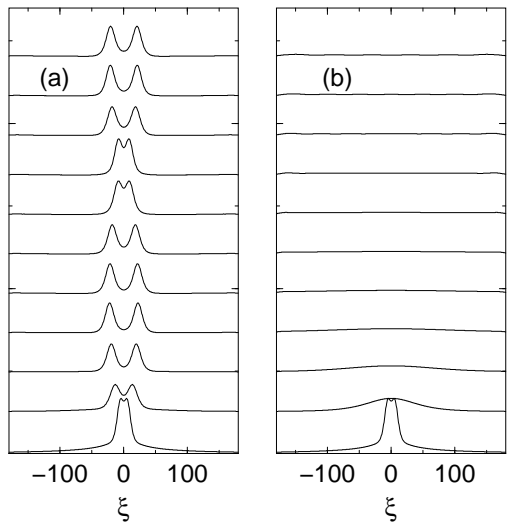}
\caption{The evolution of the system (a) with interactions
present, and (b) without interactions. In both cases, initial conditions
were simply the static HF solution from Fig. \ref{fig1},
while external potential was turned off during evolution.
Graphs show the total density profiles at equally spaced intervals
from $t=0$ up to $t=1.2\times 2\pi/\omega_x$;
the x-coordinate is $\xi=x/a_{\perp}$.
}
\label{fig2}
\end{figure}

The static evidence for mutual self trapping is confirmed
by studying the evolution of the system, once the trapping potential
along $x$ is suddenly turned off, as in the experiment of Ref. \cite{Khajkovic}.
The system is suddenly taken out of equilibrium and
consequently starts to evolve. We simulate the
dynamics by solving Eqs. (\ref{con}) and (\ref{exctu})
with the standard split-step Fourier technique.
In the spirit of Ref. \cite{Khajkovic}, we compare the $x$-unconfined
dynamics of the system in the presence of interparticle interactions
[Fig. \ref{fig2}(a)] to its time evolution when both the confinement
in $x$ and the interactions are turned off [Fig. \ref{fig2}(b)].
In the absence of interactions, we clearly observe a fast dispersion
of the total density. After approximately $t_d\sim 1/3\times 2\pi/\omega_x$,
there are hardly any particles left within the window shown in Fig. \ref{fig2}(b).
In contrast, when interactions are present, we observe self-trapped motion
[Fig. \ref{fig2}(a)]. The two humps begin to separate, because the
trapping potential which provided a balance to the kinetic energy
term is no longer present. However, due to the attractive particle
interactions, the two humps separate very slowly
and move almost in parallel [Fig. \ref{fig2}(a)].
Subsequently, the two humps are pulled back, and the initial density
roughly recovers its initial appearance.
We emphasize that a significant portion of the
atoms within the self-trapped entity ($N_1/N_c\sim 0.38$) belong
to the thermal cloud, and that they strongly affect the BEC dynamics.
Atoms from higher excited states ($j>10$) disperse quickly even
with interactions present, and are essentially spectators of the
self-trapped motion.

After the confining potential is turned off,
the condensate would be completely depleted after some
time-period, which is proportional to the collision time $\tau_{c}$,
because relaxation is attained through collisions.
The TDHF formalism is clearly inadequate to depict this
depletion since it conserves all populations including the condensate
fraction. However, since in the weak interaction regime the collisional
time is much longer than the trap period $\tau_c\omega_x\gg 1$
\cite{Kagan}, relaxation is slow compared to characteristic trap
times. We can therefore safely use the TDHF method to carry out
simulations lasting up to several dispersion times
$\tau_{disp}\sim 1/3 \times 2\pi/\omega_x\ll\tau_{c}$ in order
to demonstrate self-trapping.

\begin{figure}
\centering
\includegraphics[scale=0.6]{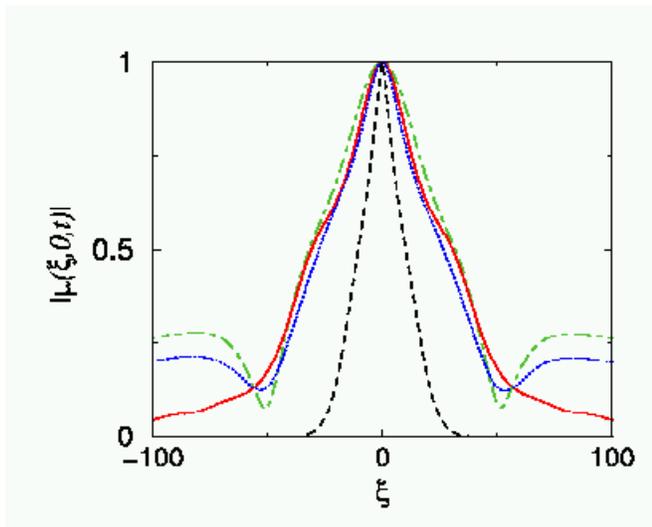}
\caption{(color online) The complex degree of coherence
$|\mu(x,x',t)|$ of a partially-coherent, self-trapped matter-wave
at times $t=0$ (black dashed line),
$t=0.4\Delta t$ (red solid line),
$t=0.8\Delta t$ (green dot-dashed line),
and $t=1.2\Delta t$ (blue dotted line);
$\xi=x/a_{\perp}$, $x'=0$, and $\Delta t=2\pi/\omega_x$.
}
\label{fig3}
\end{figure}

The self-trapped entities presented here
represent partially coherent matter waves. Thus, it is important to
study the complex degree of coherence of these matter-waves,
$\mu(x,x',t)=\rho(x,x',t)/\sqrt{\rho(x,x,t)\rho(x',x',t)}$.
Fig. \ref{fig3} shows the evolution of $|\mu(x,x',t)|$
corresponding to our self-trapped entity.
We see that spatial correlation is finite.
During the initial stage of the evolution $|\mu(x,x',t)|$
broadens indicating the increase of coherence, but after
this initial stage it stays practically unchanged.
This corresponds to the fact during the initial stage of the
evolution, a portion of the thermal cloud that
is not mutually trapped with the BEC disperses,
which initially increases the coherence.
We emphasize that for zero-temperature GPE solitons,
the pair correlation function factorizes as $\rho(x,x')=\phi^*(x)\phi(x')$,
which yields $\mu(x,x')=1$, corresponding to coherent matter-waves.
Our incoherent self-trapped matter-waves are thus rather special in that
they correspond to localization of {\it entropy} and spatial correlation,
as well as to localization of density.

Before closing, we return to the analogy between
the propagation of {\it incoherent} light in nonlinear media,
and the behavior of BECs at finite temperatures.
Incoherent light can be described by the mutual coherence
function $B(x_1,x_2,z)=\langle E^*(x_2,z,t)E(x_1,z,t)\rangle$ \cite{Equ},
where $E(x,z,t)$ is the randomly fluctuating field.
The evolution of $B(x_1,x_2,z)$ along the propagation axis $z$
is described by an equation equivalent in structure to
Eq. (\ref{exctrho}) describing the evolution of
correlations $\rho(x_1,x_2,t)$ in time (see Ref. \cite{Equ}).
Moreover, the TDHF equations (\ref{exctu}) are analogous to the Manakov
equations that describe incoherent light
in nonlinear media \cite{Equ,Mitchell1}.
The analogy is not complete, due to the fact that
the Bose wavefunction is affected by a different mean-field than the
thermal cloud, while in optics it is usual (but not the rule)
that all fields 'see' the same nonlinear change in the index of
refraction. Furthermore, in the full HFB approximation, the $U(1)$
symmetry is broken, which yields phenomena not encountered in incoherent
nonlinear optics. Nevertheless, we expect that many
nonlinear phenomena with incoherent light will find its
counterpart in BECs at finite temperatures.

In conclusion, we have demonstrated that a BEC and a portion of
its surrounding thermal cloud can exhibit motion analogous to
composite random-phase (incoherent) optical solitons.
The relation between BEC and nonlinear optics is thus elevated
to the analogy between nonlinear incoherent optical waves
and nonlinear matter-waves at finite-temperature.
The predicted incoherent matter-wave structures represent novel correlation
solitons which resemble localized second-sound entropy waves.
Work is underway to go beyond HF approximation and study the role of
pairing in lower temperature soliton dynamics.

\end{document}